\documentclass[a4paper,10pt]{scrartcl}

\usepackage{amsmath}
\usepackage{amsfonts}
\usepackage{amssymb}
\usepackage{graphicx}
\usepackage[english]{babel}
\usepackage{wasysym}
\usepackage{units}
\newcommand{\p}{\partial}
\newcommand{\reff}[1]{(\ref{#1})}
\newcommand{\vs}[1]{\vspace{#1mm}}

\newcommand{\vsO}{\vspace{.1cm}\hfill\\}
\newcommand{\vsT}{\vspace{.2cm}\hfill\\}

\title{On the Role of Viscosity in Early Cosmology}

\author{Nakia Carlevaro$^{\;a,\;b}$ and Giovanni Montani$^{\;b,\;c,\;d,\;e}$\vsT
\emph{\footnotesize $^a$Department of Physics, Polo Scientifico -- Universit\`a degli Studi di Firenze,}\vs{-2.5}\\
\emph{\footnotesize INFN -- Section of Florence, Via G. Sansone, 1 (50019), Sesto Fiorentino (FI), Italy}\\
\emph{\footnotesize $^b$ICRA -- International Center for Relativistic Astrophysics,}\vs{-2.5}\\
\emph{\footnotesize c/o Dep. of Physics - ``Sapienza'' Universit\`a di Roma}\\
\emph{\footnotesize $^c$ Department of Physics - ``Sapienza'' Universit\`a di Roma, Piazza A. Moro, 5 (00185), Rome, Italy}\\
\emph{\footnotesize $^d$ENEA -- C.R. Frascati (Department F.P.N.), Via Enrico Fermi, 45 (00044), Frascati (Rome), Italy}\\
\emph{\footnotesize $^{e}$ ICRANet -- C. C. Pescara, Piazzale della Repubblica, 10 (65100), Pescara, Italy}\vsO
{\footnotesize\ttfamily nakia.carlevaro@icra.it\quad montani@icra.it}
}
\date{}
\begin{document}
\maketitle

%
\begin{abstract} \textbf{Abstract:}
We present a discussion of the effects induced by \emph{bulk viscosity} on the very early Universe stability. The viscosity coefficient is assumed to be related to the energy density $\rho$ via a power-law of the form $\zeta=\zeta_0 \rho^s$ (where $\zeta_0,\,s=const.$) and the behavior of the \emph{density contrast} in analyzed. 

In particular, we study both Einstein and hydrodynamic equations up to first and second order in time in the so-called quasi-isotropic collapsing picture near the cosmological singularity. As a result, we get a power-law solution existing only in correspondence to a restricted domain of $\zeta_0$. The particular case of pure isotropic FRW dynamics is then analyzed and we show how the asymptotic approach to the initial singularity admits an unstable collapsing picture.
\end{abstract}

\section{The Quasi-Isotropic model}\label{Sec.2}

In 1963, E.M. Lifshitz and I.M. Khalatnikov \cite{lk63} proposed the so-called quasi-isotropic solution. This model is based on the idea that a Taylor expansion, in time, of the 3-metric can be addressed. In this approach, the Friedmann solution becomes a particular case of a larger class of solutions, existing only for space filled with matter. In fact, for an ultra-relativistic perfect fluid ($p=\rho/3$, $\rho$ being the energy density of the fluid), the spatial metric assumes the form $\gamma_{\alpha\beta}\sim a_{\alpha\beta}(x^\gamma)\,t$, asymptotically towards the singularity as $t\rightarrow0$. 

As a function of time, the 3-metric is expandable in powers of $t$. The quasi-isotropic solution is formulated in a synchronous system (\emph{i.e.}, $g_{0\alpha}=0$, $g_{00}=-1$), with a line element of the form
\begin{equation}
ds^2 = - dt^2 + \gamma_{\alpha\beta}(t, x^\gamma)dx^{\alpha}dx^{\beta}\;,\qquad
\gamma_{\alpha\beta}=t\;{a}_{\alpha\beta}+t^{2}\;{b}_{\alpha\beta}+...\;,
\end{equation}
and $a^{\alpha\beta}$ is defined as $a^{\alpha\beta}a_{\beta\gamma}=\delta_\gamma^\alpha$, moreover $b_\beta^\alpha=a^{\alpha\gamma}b_{\gamma\beta}$. Using the extrinsic curvature $k_{\alpha\beta}$, Einstein Eqs. read \cite{llfield}
\begin{equation}\nonumber
\begin{array}{ll}
R^0_0 = \tfrac{1}{2}\,\partial_t \kappa_{\alpha}^{\alpha} +
\tfrac{1}{4}\,\kappa_{\alpha}^{\beta}\kappa_{\beta}^{\alpha} = \;T^0_0-\tfrac{1}{2}T\;,
&\qquad\quad\quad
\kappa_{\alpha\beta}=\p_t\gamma_{\alpha\beta}\;,\vs{1}\\
R^0_\alpha = \tfrac{1}{2}\,(\kappa^{\beta}_{\beta ;\,\alpha}-
\kappa^{\beta}_{\alpha ;\,\beta}) = \; T_\alpha^0\;,
&\qquad\quad\quad
\kappa^\beta_\alpha=\gamma^{\beta\delta}\,\kappa_{\alpha\delta}\;,\vs{1}\\
R^\beta_\alpha = \tfrac{1}{2\sqrt{\gamma}}\,\partial_t(\sqrt{\gamma}\,
\kappa_{\alpha}^{\beta}) + P_{\alpha }^{\beta } = \;T_\alpha^\beta-\tfrac{1}{2}\,T\delta_\alpha^\beta\;,
&\qquad\quad\quad
\kappa_{\phantom{\alpha}}\;=\p_t\ln\sqrt{\gamma}\;,  
\end{array}	
\end{equation}
where $P_{\alpha }^{\beta }$ indicates the Ricci tensor and $\gamma=\det[\gamma_{\alpha\beta}]\sim t^3(1+tb)\det[a_{\alpha\beta}]$.
\vs{-2}
\paragraph{Generalized Line-Element}-\quad To include dissipative effects into the evolution of the energy source \cite{nkmIJMPD}, we deal with a more complex (no integer powers) 3-metric expansion
\begin{equation}
\gamma_{\alpha\beta}=t^{x}\;{a}_{\alpha\beta}+t^{y}\;{b}_{\alpha\beta}+...\;,\;\qquad\quad
\gamma^{\alpha\beta}=t^{-x}\;{a}^{\alpha\beta}-t^{y-2x}\;{b}^{\alpha\beta}+...\;,
\end{equation}
where we impose the constraints for the space contraction (\emph{i.e.}, $\;x>0$) and for the consistence of the perturbative scheme (\emph{i.e.}, $y>x$).
\vs{-2}
\paragraph{Viscous corrections}-\quad In this work, we treat the immediate generalization of the LK scheme considering the presence of dissipative processes within the asymptotic fluid dynamics, as expected at temperatures above $\mathcal{O}(10^{16} GeV)$. This extension is described by an additional term in the expression of the perfect fluid energy-momentum tensor:
\begin{equation}
T_{\mu\nu}= \tfrac{1}{3}\;\rho\;(4u_{\mu}u_{\nu}+g_{\mu\nu})-
\zeta\,u^{\rho}_{;\,\rho}(u_{\mu}u_{\nu}+g_{\mu\nu})\;,\qquad
u^{\rho}_{\,;\,\rho}=\p_t\ln\sqrt{\gamma}\;,
\end{equation}
where $p=\rho/3$ denotes the usual thermostatic pressure and $\zeta$ is the \emph{bulk viscosity} coefficient. In particular, here we assume this quantity as a function of the Universe energy density $\rho$; according to literature developments, we express $\zeta$ as a power-law of the form $\zeta=\zeta_0\,\rho^s$, where $\zeta_0$ is a constant and $s$ is a dimensionless parameter ($0\leqslant s\leqslant\nicefrac{1}{2}$) \cite{bk77}. In what follows, we fix the value $s=\nicefrac{1}{2}$ in order to deal with the maximum effect that bulk viscosity can have without dominating the dynamics \cite{barrow88}, in view of the phenomenological issue of thermal equilibrium perturbations which characterizes this kind of viscosity.

Writing now the \emph{00}-components of Einstein Eqs., we can expand the energy density as follows
\begin{equation}\label{pippo}
\rho=\frac{e_0}{t^2}\;+\;\frac{e_1\,b}{t^{2-y+x}}\;,\;\qquad
\sqrt{\rho}=\frac{\sqrt{e_0}}{t}\left(1\;+\;\frac{e_1\,b}{2e_0}\,t^{y-x}\right)\;,
\end{equation}
where we have assumed the condition $\;u_0^2\simeq1$\; (with $u_0=-1$) whose consistence must be verified \emph{a posteriori} and $e_0$, $e_1=const.$. 
\vs{-2}
\paragraph{The Energy Density Solution}-\quad In order to obtain an analytical expression for the energy density, we match the \emph{00}-Einstein Eq., after substituting expressions \reff{pippo}, with the hydrodynamical ones $T_{\mu;\,\nu}^{\nu}=0$. It is worth noting that, in the non-viscous case ($\zeta_0=0$), the energy density solution is determined without exploiting the hydrodynamical Eq., since $\rho$ directly comes out from the \emph{00}-gravitational Eq.. In our approximation ($u_\alpha$ is neglected with respect to $u_0$), these Eqs. can be combined together and solved order by order in the $1/t$ expansion (in the asymptotic limit $t\rightarrow0$). Since for the coherence of the solution we impose $y>x$, by solving leading-order identities we get
\begin{equation}
\label{0-sol}
x=1/[1-\tfrac{3\sqrt{3}}{4}\,\zeta_0]\;,\qquad\quad 
e_0=\tfrac{3}{4}\;x^{2}\;.
\end{equation}
The parameter $\zeta_0$ has the restriction $\zeta_0\leqslant\nicefrac{4}{3\sqrt{3}}$ in order to satisfy $x>0$: so the exponent $x$ of the metric power-law runs from $1$ (which corresponds to $\zeta_0=0$) to $\infty$. We remark that this constraint arises from a zeroth-order analysis and defines the existence of a viscous Friedmann-like model. Comparing now the two first-order identities we get
\begin{equation}
y=2\;,\;\qquad\quad 
e_1=-\tfrac{1}{2}\,x^{3}+2x^{2}-2x\;.
\end{equation}

We now narrow the validity of the parameter $x$ to the values which satisfy the constraint $x<y$, guaranteeing the consistence of the model. Thus, from \reff{0-sol}, the quasi-isotropic solution exists only if 
\begin{equation}
\label{zzero}
\zeta_0<\zeta_0^{*}=2/3\sqrt{3}\;,
\end{equation}
\emph{i.e.}, the viscosity is sufficiently small. For values of the viscous parameter $\zeta_0$ that overcome the critical one $\zeta_0^{*}$, the asymptotic quasi-isotropic solution can not be addressed, since perturbations would grow more rapidly than zeroth-order terms. 

We are now able to write the final expression of the energy density  and the density contrast ($\delta$). With the constraint $1\leqslant x<2$, we get
\begin{equation}
\rho=\frac{3\,x^{2}}{4\,t^2}\;-\;\frac{(x^{3}/2-2x^{2}+2x)\;b}{t^{x}}\;,\qquad
\delta\,=\,-\tfrac{8}{3}\,(\tfrac{1}{4}\;x+\tfrac{1}{x}-1)\,b\;t^{2-x}\;.
\end{equation}
We note that the density contrast evolution is strongly damped by the presence of dissipative effects: this behavior implies that $\delta$ approaches the singularity more weakly as $t\rightarrow0$ when the viscosity runs to $\zeta_0^{*}$. In correspondence with this threshold value the density contrast remains constant in time and hence it must be excluded by the possible $\zeta_0$ choices.      

We conclude this section verifying the consistence of our model. By the analysis of $\alpha$\emph{0}-gravitational Eq. we get, up to the dominant-order of expansion: $u_\alpha\sim\;t^{3-x}$. The assumption $u_0^{2}\simeq1$ is, therefore, well verified: in fact, $u_\alpha u^\beta\sim t^{6-3x}$ can be neglected in the contraction $u_\mu u^\mu=-1$ and our approximation scheme results to be self-consistent.

\section{The Pure Isotropic FRW Model}
In this Section, we briefly investigate the effects that bulk viscosity has on the stability of the pure isotropic Friedmann-Robertson-Walker (FRW) Universe \cite{nkmFRW,padbanabham}. We start here from the perturbed Einstein Eqs. in a FRW background filled with ultra-relativistic viscous matter, whose coefficient $\zeta=\zeta_0\,\rho^{s}$ corresponds to the choice $s=\nicefrac{1}{2}$. The zeroth-order dynamics is described by the energy conservation Eq. and the Friedmann one. Assuming the conformal time $\eta$ ($dt=ad\eta$), the solutions are
\begin{equation}
\label{omega}
\rho=C a^{-(2+2\omega)}\;,\quad 
a=a_1\,\eta^{1/\omega}\;,\quad
\omega=1-\chi\,\zeta_0\,,\quad
\chi=\sqrt{54\pi G}\;,
\end{equation}
being $C$ an integration constant and $a_1=(8\omega^2\pi C G/3)^{1/2\omega}$. Since we consider an expanding Universe, we have $\omega>0$ obtaining the constraint $0\leqslant\zeta_0<\nicefrac{1}{\chi}$.

The analysis of perturbations moves from the viscous energy-momentum tensor and the scalar representation of the perturbed \emph{3}-metric $h_{\alpha}^{\beta}$ in a synchronous reference system \cite{llfield}. Solving the perturbed Einstein Eq. we obtain the following expression for the the density contrast $\delta=\nicefrac{\delta\rho}{\rho}$
\begin{equation}
\label{contr-visc}
\delta\sim[C_1\eta^{3-\nicefrac{2}{\omega}}+C_2\eta^2+C_3\eta^{3-\nicefrac{1}{\omega}}
+C_4\eta^{5-\nicefrac{1}{\omega}}],
\end{equation}
where $C_{1,2,3,4}$ are constants.

As issue of this analysis, we find that two different dynamical regimes appear when viscosity is taken into account and the transition from one regime to the other one takes place when the parameter $\zeta_0$ overcomes the given threshold value $\zeta_0^{*(FRW)}=\nicefrac{1}{3\chi}$. However, in both these stages of evolution, the Universe results to be stable as it expands; the effect of increasing viscosity is that the density contrast begins to decrease with increasing time when $\zeta_0$ is over the threshold. It follows that a real new feature arises with respect to the non-viscous analysis when the collapsing point of view is addressed: if $\zeta_0>\zeta_0^{*(FRW)}$, the density contrast explodes asymptotically and the isotropic Universe results unstable approaching the initial singularity. 

We conclude comparing the pure isotropic picture to the quasi-isotropic one since also the pure Friedmann-singularity scheme is preserved only if we deal with limited values of the viscous parameter. The threshold values of the the two pictures satisfy the condition $\zeta_0^{*(FRW)}<\zeta_0^{*}/3$ (see Eq. \reff{zzero} using geometric units): this constraint is physically motivated if we consider, as it is,  the FRW model as a particular case of the quasi-isotropic solution.

\end{document}